\documentclass[12pt]{iopart}

\usepackage{iopams}  
\usepackage{graphicx}

\begin{document}

\title[]{Simplifying higher-order perturbation theory for ring-shaped Bose-Hubbard systems}

\author{Meret Preuß}

\address{Carl von Ossietzky Universität Oldenburg,\\
	Ammerländer Heerstraße 114-118,\\
	26129 Oldenburg}
\ead{meret.preuss@uni-oldenburg.de}
\vspace{10pt}
\begin{indented}
\item[]\today
\end{indented}

\begin{abstract}
In this paper, higher-order perturbation theory is applied and tailored to one-dimensional ring-shaped Bose-Hubbard systems. Spectral and geometrical properties are used to structurally simplify the contributions and reduce computational effort without sacrificing accuracy. For this, a guide for the computation of the individual perturbational orders up to order nine is provided, alongside a both system-specific and parametrization-dependent convergence criterion. The simplification scheme described is found to be applicable to a wider class of Bose-Hubbard systems with different lattice geometries. An exemplary validation of these findings is included in the form of explicit calculations of ground state energies of the three-site Bose-Hubbard system with repulsive on-site interactions. These calculations are successfully checked against numerical computations of exact diagonalization results.
\end{abstract}

%
\vspace{2pc}

%
\submitto{\PS}
%
%
%

\section{Introduction}
The Bose-Hubbard model \cite{gersch1963quantumcellmodel} describes spinless bosons that interact on a lattice. Despite its rather minimalistic form constisting solely of a tunneling part and on-site interactions, this system allows for an exploration of quantum phase transitions \cite{jaksch1998coldbosonicatoms, freericks1994phasediagrambosehubbard} and lattice dynamics \cite{elstner1999dynamicsthermodynamicsbosehubbard}. Moreover, more complicated systems can be derived from the original Bose-Hubbard model as shown in \cite{penna2017two} with two different condensates or in \cite{richaud2017quantum} in the context of a two-ring ladder.

In the recent decades, Bose-Hubbard systems with different lattice geometries have been intensely studied using various methods like numerics \cite{bhattacharyya2023operatorgrowthkrylov}, semiclassical analysis \cite{arwas2015superfluiditychaoslow,arwas2014triangularbosehubbardtrimer} algebra  \cite{pan2014symmetrybasedapproachgroundstate,pan2023exactsolutionhomogenous}, random matrix theory \cite{kolovsky2016bosehubbardhamiltonian} and perturbational methods for one-  \cite{damski2006mottinsulatorphaseonedimensional} and higher-dimensional lattices \cite{hinrichs2013perturbativecalculationcritical, wang2018high} - the latter being the method of choice for this paper, as well. In general, time-independent perturbation theory, classically known as \textit{Rayleigh-Schrödinger} perturbation theory \cite{rayleigh1896theory,schrodinger1926quantisierungalseigenwertproblem} contains a well established and proven framework for perturbative calculations. However, its application poses the danger of only mechanically applying the key formalism without gaining insight into the physics behind it. In contrast to the recursive calculations of the series' coefficients, Kato \cite{kato1949convergenceperturbationmethod} proposed a method that allows for an individual consideration of the separate orders while deriving the mathematically identical perturbational power series as Rayleigh and Schördinger did. Yet, for higher orders, the super-exponentially growing number of terms can become rather extensive to compute. However, as shown by Eckardt \cite{eckardt2009processchainapproachhighorder} this amount of formally derived terms can be reduced by a priori including information about the considered system - for instance, if the first order correction is known to vanish. Works making use of this approach are for instance \cite{heil2012strong} and \cite{sanders2019quantum}. In this current paper, that approach is extended and applied to one-dimensional Bose-Hubbard systems with a ring-like lattice geometry whose parametrization lies within the Mott regime.

 In an effort to render the structural understanding of perturbation theory more accessible, spectral and geometrical properties of Bose-Hubbard systems are used to systematically investigate and simplify the individual contributions to the perturbational series, thus aiming at reducing the number of terms that explicitly need to be calculated to obtain the $n$-th order correction. 
 For this, Section \ref{sec:BH} introduces ring-shaped Bose-Hubbard systems alongside their spectral properties, while Section \ref{sec:Kato} provides an overview of the perturbation theory as formulated by T. Kato \cite{kato1949convergenceperturbationmethod,kato2013perturbationtheorylinear}. Section \ref{sec:KatoBH} then contains the application and simplification of said theory to the described Bose-Hubbard systems, as well as an inspection of convergence properties and explicit perturbative calculations of ground state energies.

\section{Bose-Hubbard Rings}\label{sec:BH}
The systems considered in this paper are one-dimensional ring-shaped Bose-Hubbard models (which in the following shall be referred to as \textit{Bose-Hubbard rings}) \cite{amico2005quantummanyparticle} which describe $N$ spinless bosons interacting on a lattice. The respective dimensionless Hamiltonian operator $\widehat{H}$ for a ring lattice with $M$ sites is expressed through the bosonic creation and annihilation operators $\hat{a}^\dagger_i, \hat{a}_i$ ($i=1,...,M$) obeying $[\hat{a}_i, \hat{a}^\dagger_j] = \delta_{ij}$. $\widehat{H}$ is parameterized by the ratio of the particles' non-linear interaction strength $\kappa$ and tunneling frequency $\Omega$.
\begin{eqnarray}
\widehat{H} = - \frac{\Omega}{\kappa} \sum\limits_{i,j = 1 \atop i \neq j}^{M} 
\hat{a}^\dagger_i \hat{a}_j + \frac{1}{2} \sum\limits_{i = 1}^{M} \hat{a}^\dagger_i \hat{a}^\dagger_i \hat{a}_i \hat{a}_i \label{eq:Hamiltonian}
\end{eqnarray}
The kinetic part of $\widehat{H}$, i.e. the tunneling term, will here be considered a small perturbation in comparison to the potential energy term which is governed by the particles' on-site interactions. The latter can also be expressed using the operators $\hat{n}_i$ for the number of particles $n_i$ at the $i$-th lattice site. As $[\widehat{H}, \hat{n}] = 0$,  both $\widehat{H}$ and $\hat{n}$ share a common system of eigenfunctions. The eigenstates of $\hat{n}$ are called Fock states and span the $N$-particle sector of the Fock space $\mathcal{H}_N$. \cite{bhattacharyya2023operatorgrowthkrylov,pan2023exactsolutionhomogenous}
As the potential energy term is already diagonal in Fock representation, it will later be considered the unperturbed part $\widehat{H}_0$ in the perturbative analysis.
\begin{eqnarray}
	\widehat{H}_0 = \frac{1}{2} \sum_{i=1}^M \hat{n}_i (\hat{n}_i-1)
\end{eqnarray}
$H_0$, which is governed by $\sum_{i=1}^M \widehat{n}_i^2$, becomes minimal for an equal distribution of the $N$ particles onto the $M$ lattice sites, i.e. $n_i = N/M~, i = 1...M$.  \footnote{This only applies to systems with repulsive on-site interactions. For attractive systems (see \cite{buonsante2005attractiveultracoldbosons}) for which the unperturbed part would be $-\hat{H}_0$, the ground state is $M$-fold degenerate and characterized by the localization of all $N$ particles at the same lattice site. In this case, degenerate perturbation theory needs to be applied.} For simplicity, this paper will consider $N$ to be a multiple of $M$.  Using this equal distribution, the ground state energy of the potential energy Hamiltonian of the  $M$-site system can be calculated as
\begin{eqnarray}
	E_{\mathrm{0,GS}} = \frac{1}{2} \left(\frac{N^2}{M}-N\right).
\end{eqnarray}
The respective first excited states are formed by taking the ground state configuration and moving one boson to another lattice site, i.e. for instance 
\begin{eqnarray}
	n_i = \cases{\frac{N}{M} \qquad ~\ i = 3...M,\\ \frac{N}{M} + 1 \quad i = 1,\\ \frac{N}{M} -1\quad i = 2. }
\end{eqnarray}
Due to the system's symmetry, this state is $M(M-1)$-fold degenerate. Its energy eigenvalue can be calculated to be 
\begin{eqnarray} E_{\mathrm{0,ex}}&= \frac{1}{2} \left(M-2\right)\left(\frac{N^2}{M^2}-\frac{N}{M}\right) + \left(\frac{N}{M}-1\right)^2 \nonumber \\ & \quad-\left(\frac{N}{M}-1\right)+\left(\frac{N}{M}+1\right)^2 - \left(\frac{N}{M}+1\right ) \nonumber\\ &= \frac{1}{2}\left(\frac{N^2}{M}-N \right)+1 .
\end{eqnarray}
Thus, the minimum energy distance $d$ between the unperturbed system's ground state and the rest of the spectrum is always one. \footnote{With physical units, i.e. a dimensional Hamiltonian, this spectral gap is $1~\hbar \kappa$~.} This finding will later allow us to greatly simplify the perturbational series for Bose-Hubbard rings of arbitrary size $M$. 
As a minimal example for such a Bose-Hubbard ring, the three-site system (also called the Bose-Hubbard trimer) will serve as a model system. Thus, for the trimer, the ground state of $\widehat{H}_0$ is characterized by $n_i = N/3 ,~ i = 1,2,3$. \ref{sec:appendix} provides the explicit form of the trimer Hamiltonian in matrix representation, as well as a description of the method used to label and sort the individual Fock states.

\section{Perturbation Theory by T. Kato}\label{sec:Kato}
For a general Hamiltonian $\widehat{H} = \widehat{H}_0 + \lambda \widehat{V}$ composed of a diagonalized part $\widehat{H}_0$ and a perturbation $\widehat{V}$ scaled by ${\lambda\in\mathbb{R}}$,  the perturbational series for the energy eigenvalue $E_a$ (originating from the non-perturbed and non-degenerate eigenvalue $\varepsilon_a$ of an eigenstate $|a\rangle$), as  formulated first by T. Kato \cite{kato2013perturbationtheorylinear}, is given by:
\begin{eqnarray}
	E_a = \varepsilon_a + \sum_{n=1}^\infty \lambda^n~ \Tr\widehat{B}^{(n)},
\end{eqnarray} 
with
\begin{eqnarray}
	\widehat{B}^{(n)} =  \sum_{\sum_{i=1}^{n+1} k_i = n-1} \widehat{S}^{k_1}\widehat{V}\widehat{S}^{k_2}\widehat{V}...\widehat{V}\widehat{S}^{k_n+1}~.
\end{eqnarray}
The operators $\widehat{B}^{(n)}$ are defined using the projectors $\widehat{S}^k$ given by 
\begin{eqnarray}
	\widehat{S}^k = \cases{-|a\rangle\langle a| \qquad \quad \quad \mathrm{for}~ k = 0 \\
		\sum_{i \neq a} \frac{P_i}{\left (\varepsilon_a-\varepsilon_i \right )^k} \quad \,~ k \geq 1 .}
\end{eqnarray}
with $P_i$ being the projector onto the eigenspace of an unperturbed eigenvalue $\varepsilon_i$.  A detailed derivation of the above-mentioned findings is given, for instance, in \cite{kato2013perturbationtheorylinear}. 
As described in Ref, \cite{eckardt2009processchainapproachhighorder}, in view of the operators $\widehat{B}^{(n)}$, the determination of the $n$-th order correction consists of finding all possible ($n+1$)-element sets of $k_i$ which each sum up to $n-1$. Besides, the combinatorial condition ensures the presence of at least two $k_i$ being zero within each set. With this, cyclic permutations (leaving the trace unchanged) together with the properties $\widehat{S}^0 = -\left (\widehat{S}^0 \right )^2$, $\widehat{S}^{k_1} \widehat{S}^{k_2} = \widehat{S}^{k_1+k_2}$ help to convert the trace into a diagonal matrix element (DME) with respect to the unperturbed state \footnote{if using matrix representations of operators, as done here.}. For this consider one explicit $\widehat{S}^0$ within $\widehat{B}^{(n)}$ and $C$ and $D$ as the corresponding ``remainders''. The following then holds: 
\begin{eqnarray}
	\Tr(C\widehat{S}^0D) & = \Tr(\widehat{S}^0CD)\\
	& = \Tr(-\widehat{S}^0\widehat{S}^0CD)\\
	& = -\Tr(|a\rangle \langle a |CD|a\rangle \langle a| )\\
	& =  - \langle a |CD|a\rangle \end{eqnarray}
The remaining operator $CD$ now contains an $n$-element set of $k_i$ that add up to $n-1$ - including, still, at least one $k_i$ being zero. An explicit writing of $\widehat{S}^0$ then splits the DME into a product of two DMEs. If a DME does not contain any occurrence of $\widehat{S}^0$, it will now be referred to as an elementary matrix element (EME). For the sake of brevity, each summand within $\widehat{B}^{(n)}$ will henceforth be represented by products of its EMEs which themselves are denoted by their (sub-)sets of exponents $k_i$. For instance, the fourth order term \footnote{There is no minus sign, since here, the explicit writing of the outer $\hat{S}^0$ directly delivers the form of a DME.}
\begin{eqnarray}
	\mathrm{Tr}\left(\widehat{S}^0\widehat{V}\widehat{S}^2\widehat{V}\widehat{S}^0\widehat{V}\widehat{S}^1\widehat{V}\widehat{S}^0\right)= \langle a |\widehat{V}\widehat{S}^2\widehat{V}\widehat{S}^0\widehat{V}\widehat{S}^1\widehat{V}|a\rangle
\end{eqnarray}
is denoted by $(2,0,1)\equiv - (2)(1)$.  With this, each DME can be represented (and calculated) by its ``occupation'' of the individual EMEs. \footnote{For a more thorough deviation and further examples, the reader is referred to \cite{eckardt2009processchainapproachhighorder}.} The factorization of a DME into EMEs allows for an identification of formally different DMEs which, owed to the products' commutativity, are identical in value. With this, the DMEs can then be grouped into ``families'' that each are assigned a weight factor arising from the combination of the individual DMEs and their sign.  
Thus, each order of energy correction $n$ is represented by a collection of DME families, expressed by their EME occupation, and the corresponding weight factors. 
Considering the rather abstract results of the formalism introduced by Kato, it is important to note that the respective generated power series of the control parameter $\lambda$ are exactly the same ones that one obtains by applying the formalism by Rayleigh and Schrödinger \cite{kato2013perturbationtheorylinear}.\\
A Python-implementation of the Kato formalism described above (for non-degenerate eigenvalues) for any quantum-mechanical system with a finite-dimensional Hamiltonian $\widehat{H}$ is available \cite{zenodo_15005363}. \footnote{In the code, all steps are independent of any particular system and only need to be performed once. In addition, the code contains scripts for an explicit application of perturbation theory for a two-level system and the Bose-Hubbard trimer.}

\section{Perturbation theory for Bose-Hubbard rings}\label{sec:KatoBH}
As known from (particularly) Rayleigh-Schrödinger perturbation theory, the first order energy correction is given by $\langle a|\widehat{V}|a\rangle$, i.e. the diagonal matrix element of the perturbation operator $\widehat{V}$ with respect to the unperturbed state $|a\rangle$. If, however, $\widehat{V}$ does not possess any diagonal elements in the eigenbasis of $\widehat{H}_0$, this first order correction is zero. Looking at the factorization of the diagonal matrix elements into elementary matrix elements, each term containing $(~)$ can be discarded. This allows for a notable reduction of the number of terms in the perturbational series - especially in higher orders, as shown in \cite{eckardt2009processchainapproachhighorder}. 
\subsection{Structural simplification of the contributions}\label{sec:simpli}
In the following, the Kato formalism will be applied to the ground state of Bose-Hubbard rings. In addition to the removal of all terms containing a first order correction, the known spectral gap between the ground state and the first excited states allows for a structural simplification and association of differents EMEs with one another. In the following, thus adopting the symbolics explained in Section \ref{sec:Kato}, EMEs with the representation $(...,k_i)$ symbolize the matrix elements $\langle a| ... \widehat{S}^k \widehat{V} |a\rangle$. The perturbation operator $\widehat{V}$ only contains nearest neighbor-couplings. Working on the ground state, $\widehat{V} |a\rangle$ is a superposition of all first excited states that are accessible from the ground state within a single tunneling event - all of which are associated with the same spectral gap $\varepsilon_a-\varepsilon_i = -1$ (with $i\neq a$). Thus, together with the projectors' $P_i$ idempotence, one finds that
\begin{eqnarray} \widehat{S}^k \widehat{V} |a\rangle &= (-1)^{k-1} \widehat{S}^1 \widehat{V} |a\rangle. 
\end{eqnarray}
Due to the hermiticity of $\widehat{V}$ and all $\widehat{S}^k$, all $m$-digit EMEs of the form $(k_1,k_2,..., k_{m-1},k_m)$ give the same contribution as their reflected versions $(k_m, k_{m-1},...k_2,k_1)$. The combination of the ``reduction'' of arbitrary powers $k$ of $\widehat{S}^k$ ($k \geq 1$) and the EMEs' reflection symmetry now lead to several explicit key findings for the evaluation of the contributing terms in the perturbational series of Bose-Hubbard rings:\\
\begin{enumerate}
	\item All one-digit EMEs $(k)$ can be written as $(-1)^{k-1} (1)$.
	\item All two-digit EMEs $(k_1,k_2)$ can be written as $(-1)^{k_1+k_2-2}(1,1)$.
	\item Three-digit EMEs with the central exponent $k_c$  $(k_1, k_c,k_2)$ can be written as $(-1)^{k_1+k_2-2}(1,k_c,1)$.
	\item All further $m$-digit EMEs $(k_1,k_2,..., k_{m-1},k_m)$ can be simplified to \\$(-1)^{k_1+k_m-2}(1,k_2,..., k_{m-1},1)$, along with possible combinations due to reflection symmetries. 
\end{enumerate}
These four findings allow for a substantial reduction of the number of elementary terms that need to be explicitly calculated. \tref{tab:DMEs} contains the results of the simplifications described above up to order $n=9$. These general findings are not only applicable to all Bose-Hubbard rings, but to all Bose-Hubbard systems consisting of a hermitian perturbation with only one tunneling event $\widehat{V}$ and on-site interaction term as contained in \eref{eq:Hamiltonian}. However, in this paper, further focus will only be laid upon Bose-Hubbard rings. Assuming convergence (see more on this in Section \ref{sec:convergence}), the different system-parameter-specific perturbational series are obtained by determining the explicit values of the elementary matrix elements and summing up all weighted DMEs in each order. \Tref{tab:numcontributions} contains supplementary information on the scale of reduction of the contributing terms that is possible with the help of the EME simplification - and thus information on the computational effort needed to explicitly determine the perturbational series up to order $n=9$.

\fulltable{\label{tab:DMEs}Remaining (weighted) diagonal matrix elements (written as products of elementary matrix elements) after the structural simplification applicable to the perturbation series for Bose-Hubbard systems up to order $n=9$. A summation of all weighted terms in one order yields the system-specific coefficient of $\left(\Omega/\kappa\right)^n$ in the perturbational series when explicitly calculating the values of the elementary matrix elements.}
	\begin{tabular}{@{}rrrrr@{}}\br
		$n$ & DMEs                  &                        &                        &                        \\ \cr \mr
		1   & -                     &                        &                        &                        \\ \cr  \mr
		2   & $(1)$                 &                        &                        &                        \\ \cr \mr
		3   & $(1,1)$               &                        &                        &                        \\ \cr \mr
		4   & $(1)^2$               & $+ (1,1,1)$            &                        &                        \\ \cr \mr
		5   & $3(1)(1,1)$           & $+(1,1,1,1)$           &                        &                        \\ \cr \mr
		6   & $2(1)^3$              & $ +3 (1)(1,1,1)$       & $ + (1,1,1,1)$         & $-(1)(1,2,1)$          \\
		& $+2(1,1)^2$           &                        &                        &                        \\ \cr \mr
		7   & $11(1)^2(1,1)$        & $+3(1)(1,1,1,1)$       & $+4(1,1)(1,1,1)$       & $-2(1)(1,1,2,1)$       \\
		& $-(1,1)(1,2,1)$        & $+(1,1,1,1,1,1)$       &                        &                        \\ \cr \mr
		8   & $5(1)^4$              & $ + 11(1)^2(1,1,1)$    & $-6(1)^2(1,2,1)$       & $ +(1)^2(1,3,1)$       \\
		& $ + 17(1)(1,1)^2$     & $-2(1)(1,1,1,2,1)$     & $-(1)(1,1,2,1,1)$      & $+4(1,1)(1,1,1,1)$     \\
		& $ -2(1,1)(1,1,2,1)$   & $+2(1,1,1)^2$          & $-(1,1,1)(1,2,1)$      & $+3(1)(1,1,1,1,1)$     \\
		& $+(1,1,1,1,1,1,1)$    &                        &                        &                        \\ \cr \mr
		9   & $40(1)^3(1,1)$        & $ + 11(1)^2(1,1,1,1)$  & $ - 12(1)^2(1,1,2,1)$  & $ + 2(1)^2(1,1,3,1)$   \\
		& $2(1)^2(1,2,2,1)$     & $+ 34 (1)(1,1)(1,1,1)$ & $ -14 (1)(1,1)(1,2,1)$ & $ + 2(1)(1,1)(1,3,1)$  \\
		& $+3(1)(1,1,1,1,1,1)$  & $-2(1)(1,1,1,1,2,1)$   & $-2(1)(1,1,1,2,1,1)$   & $+ 8(1,1)^3$           \\
		& $ +2(1,1)(1,1,1,1,1)$ & $ -2(1,1)(1,1,1,2,1)$  & $ -1(1,1)(1,1,2,1,1)$  & $ + 4(1,1,1)(1,1,1,1)$ \\
		& $ -2(1,1,1)(1,1,2,1)$ & $-(1,1,1,1)(1,2,1)$    & $+ 2(1,1)(1,1,1,1,1)$  & $ + (1,1,1,1,1,1,1,1)$ \\ \cr \br
	\end{tabular}
\endfulltable

\begin{table*}[]
	\caption{Number of quantities that explicitly need to be calculated for an order $n$ of perturbation theory. The \textit{left} columns show findings from the general (system-unspecific) Kato Formalism: number of different diagonal matrix elements (DMEs) with / without a nonzero first-order correction, as well as the number of elementary matrix elements that need to be calculated as ``building blocks'' for a certain order. The second number for the latter indicates how many of those EMEs have not been calculated (and thus possibly stored) in the previous orders. The \textit{right} columns contain numbers for the simplified Kato perturbation series as obtained using the rules in section \ref{sec:simpli}): The number of DMEs in the order $n$, as well as the number of (reduced) EMEs that need to be calculated / that have not appeared in the previous orders. The last row contains the respective sum of all orders above, i.e. it gives information about the computational effort needed to calculate the first nine orders of the perturbational series.}\label{tab:numcontributions}
	\begin{indented}\item[]
		\begin{tabular}{ccclcc@{}}\br
			& \multicolumn{2}{c}{Standard Kato Formalism}                      &  & \multicolumn{2}{c}{Simplification for Bose-Hubbard systems} \\ \cr \mr
			$n$ & \#DMEs & \#needed EMEs / new &  & \#DMEs & \#needed EMEs / new   \\ \cr \mr
			\multicolumn{1}{c|}{1}   & 1/0                                   & 0/0                        &  & 0                           & 0/0                         \\
			\multicolumn{1}{c|}{2}   & 1/1                                   & 1/1                        &  & 1                           & 1/1                         \\
			\multicolumn{1}{c|}{3}   & 2/1                                   & 1/1                        &  & 1                           & 1/1                         \\
			\multicolumn{1}{c|}{4}   & 4/2                                   & 3/2                        &  & 2                           & 2/1                         \\
			\multicolumn{1}{c|}{5}   & 10/3                                  & 5/2                        &  & 2                           & 3/1                         \\
			\multicolumn{1}{c|}{6}   & 22/7                                  & 9/4                        &  & 5                           & 5/2                         \\
			\multicolumn{1}{c|}{7}   & 53/12                                 & 14/5                       &  & 6                           & 7/2                         \\
			\multicolumn{1}{c|}{8}   & 119/26                                & 23/9                       &  & 13                          & 11/4                        \\
			\multicolumn{1}{c|}{9}   & 278/47                                & 35/14                      &  & 20                          & 16/6 \\ \cr \mr
			\multicolumn{1}{c|}{$\sum$}   & 490/99                                & 91/38                      &  & 50                          & 46/18 \\ \cr \br                   
		\end{tabular}
		\end{indented}
\end{table*}
\subsection{Convergence}\label{sec:convergence}
For regular perturbations $\widehat{V}$ to hermitian operators $\widehat{H}_0$, \cite{kato1949convergenceperturbationmethod} gives an estimation on the convergence radius for the perturbational series. With $d$ being the so-called distance of isolation of the eigenvalue $\varepsilon_0$ (to the rest of the spectrum of $\widehat{H}_0$), the series is estimated to absolutely converge if the following holds for the control parameter (here: $\Omega/\kappa$):
\begin{eqnarray}
	\left | \frac{\Omega}{\kappa}\right | \left | \left | \widehat{V} \right | \right | \leq \frac{d}{2} \label{eq:conv}
\end{eqnarray}
Here, the factor $1/2$ is a conservative statement - Kato calls the criterion in \eref{eq:conv} the ``crudest estimate without considering special properties of [$\widehat{V}$]''. In his book, \textit{Perturbation Theory for Linear Operators}, Kato \cite{kato2013perturbationtheorylinear} specifies the used norm $|| \widehat{V} ||$ to be the spectral norm. With $x$ being a vector within the space of $\widehat{V}$, this norm is defined as
\begin{eqnarray}
	||\widehat{V}|| &= \sup_{||x||=1}||\widehat{V}x|| .
\end{eqnarray}  
However, at the same time, this operator norm is given by the maximum singular value $\sigma_{\mathrm{max}}$. In case of a hermitian operators $\widehat{V}$, this is equivalent to the largest absolute eigenvalue $|\varepsilon_{\mathrm{max}}|$ \cite{horn2012matrixanalysis}. 
\begin{eqnarray}
	||\widehat{V}||& = \sigma_{\mathrm{max}} \\
	& = |\varepsilon_{\mathrm{max}}|
\end{eqnarray} 
For one-dimensional Bose-Hubbard-Hamiltonians, the spectrum of the tunneling operator is best studied in the Bloch basis. For this, \cite{kolovsky2016bosehubbardhamiltonian} gives an expression for the energy eigenvalues of $\widehat{V}$:
\begin{eqnarray}
	E_V = -2 \sum_k \cos\left (\frac{2\pi k}{M}\right )n_k
\end{eqnarray}
Here, $n_k$ describes the number of bosons in the Bloch state with quasimomentum $2 \pi k/M$. $E_V$ is maximal if all bosons occupy the Bloch state specified by $k=0$. Thus, the operator norm is independent of the number of lattice sites $M$ and is given by
\begin{eqnarray}
	\left |\lambda_{\mathrm{max}}\right | = \left | \left | \widehat{V} \right | \right | = 2N.
\end{eqnarray}
Combining the knowledge on $\left | \left | V \right | \right |$ and the distance of isolation ($d=1$), the size of the radius of convergence (as given by \eref{eq:conv}) for the Bose-Hubbard ring, without any dependence on the number of lattice sites $M$, can be expressed as being inversely proportional to the system size $N$.
\begin{eqnarray}
	\left | \frac{\Omega}{\kappa}\right | \leq \frac{1}{4N}\label{eq:convBH}
\end{eqnarray}
\subsection{Calculations}\label{sec:calc}
\begin{figure*}[]
	\centering
	\includegraphics[width = \textwidth]{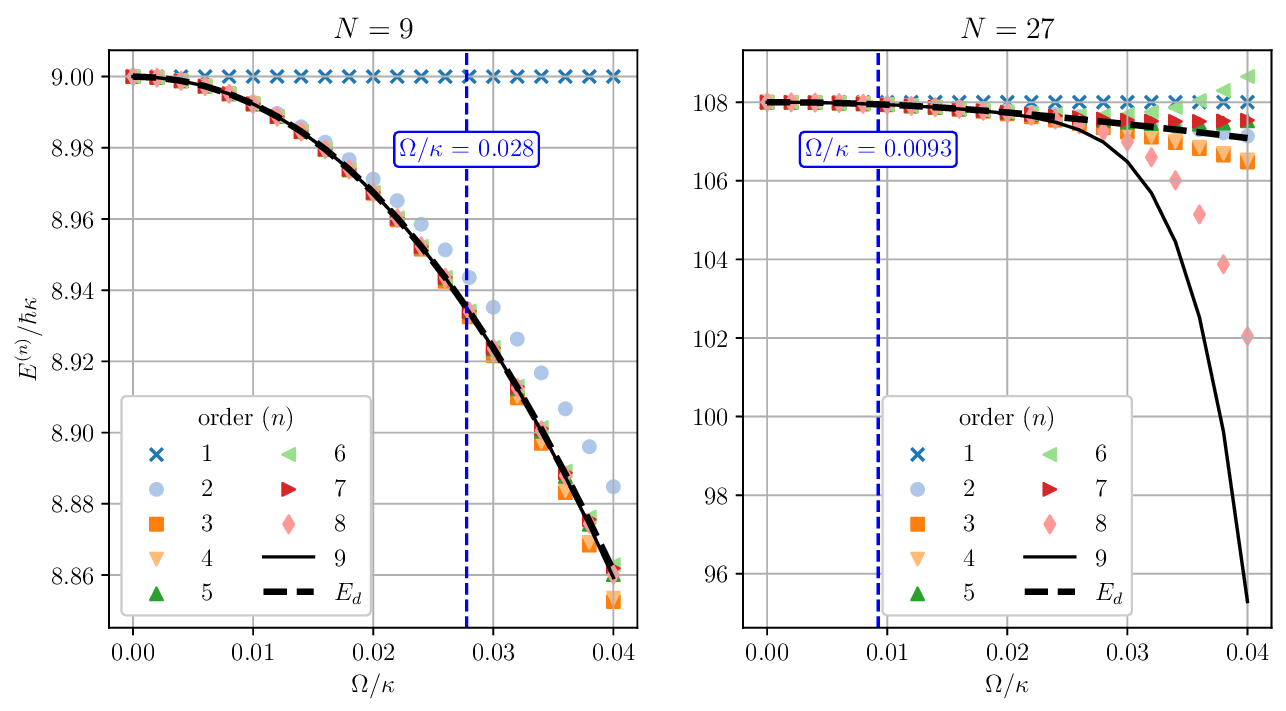}
	\caption{Total energy corrections in units of $\hbar \kappa$ for the ground state energies of Bose-Hubbard trimers with $N = 9/27$ and  different parametrizations $\Omega/\kappa$ for perturbational series up to different orders $n$, as well as the results from numerical diagonalization $E_d$. Convergence radii calculated using \eref{eq:convBH} are given as vertical dashed lines.}\label{fig:conv927}
\end{figure*}

\begin{figure*}[]
	\centering
	\includegraphics[width = \textwidth]{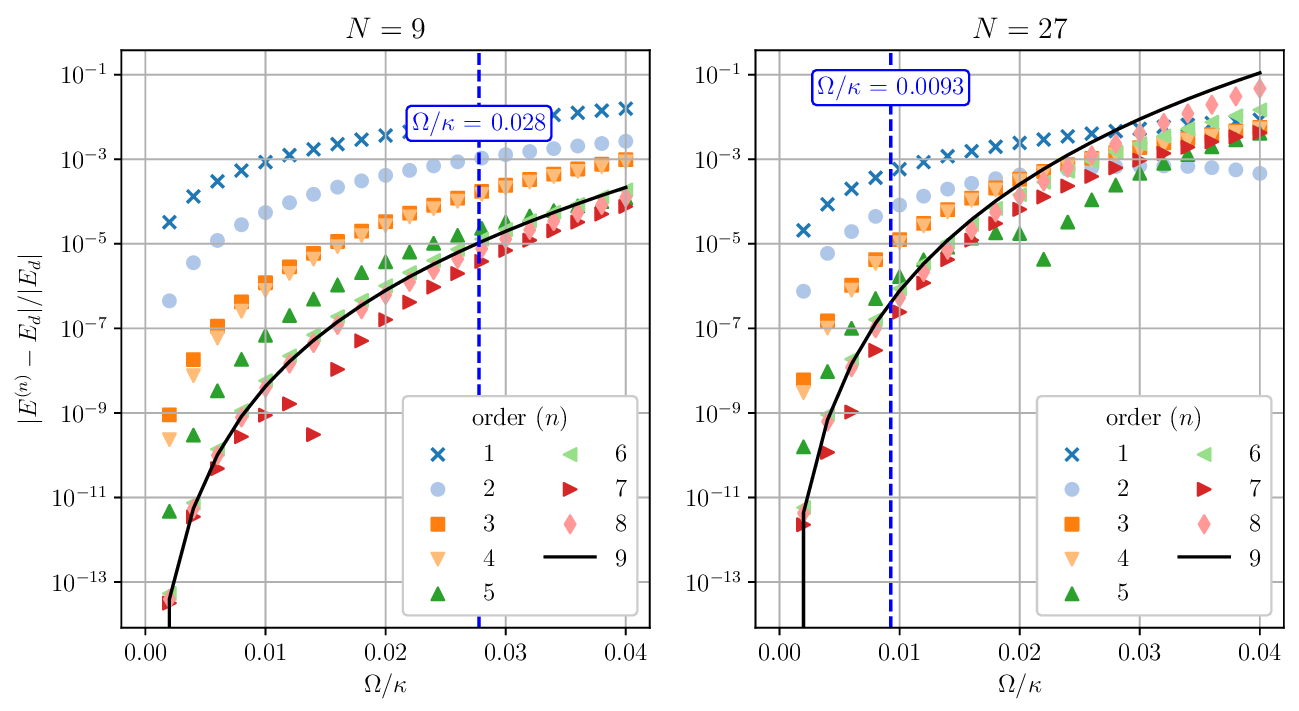}
	\caption{Relative energy deviations of the $n$-th order perturbative calculations $E^{(n)}$ to the numerically calculated result $E_d$ using a full diagonalization procedure for system sizes $N=9/27$ of Bose-Hubbard trimers with different parametrizations $\Omega/\kappa$ (logarithmic scale). The convergence radii, as computed with \eref{eq:convBH}, are indicated by vertical dashed lines. The highest calculated order ($n=9$) is shown with a solid line.}\label{fig:conv927rel}
\end{figure*}

\begin{figure*}[]
	\centering
	\includegraphics[width = \textwidth]{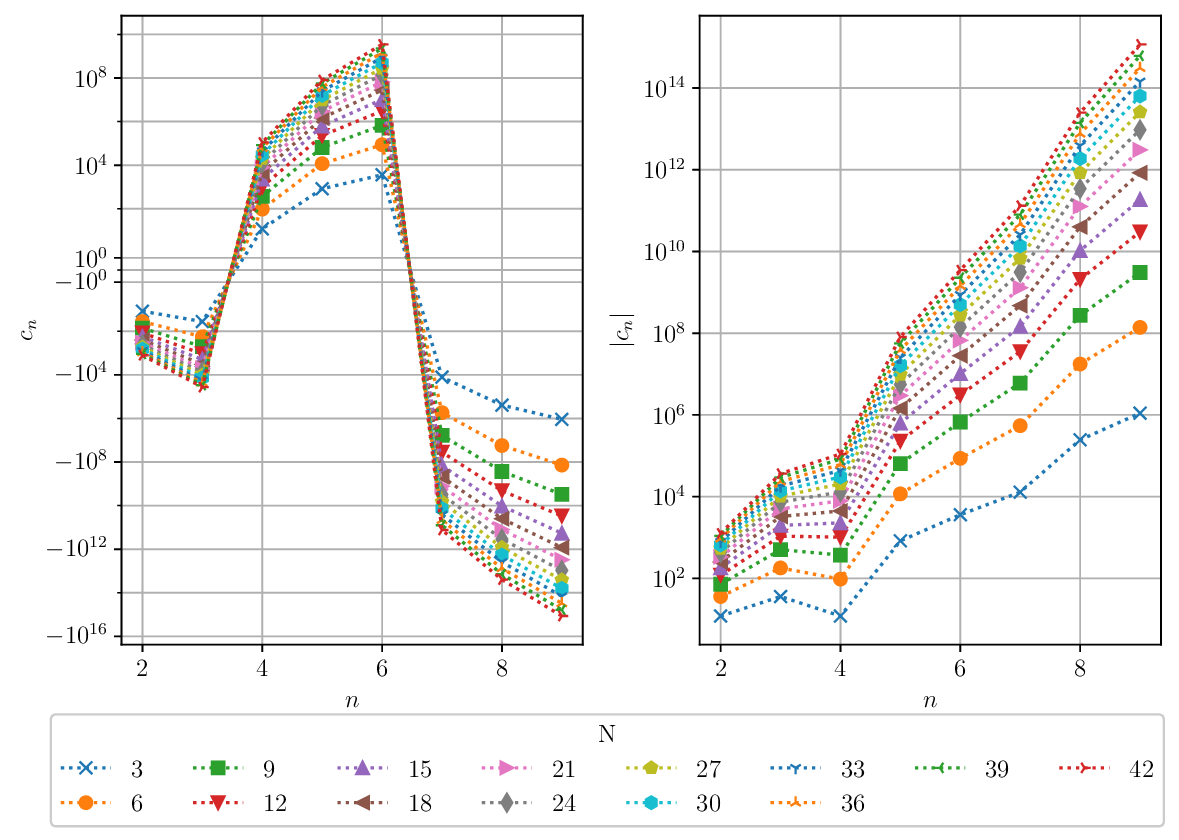}
	\caption{$n$-th order coefficients $c_n$ of the perturbative calculations for the Bose-Hubbard trimer with different system sizes $N$ plotted by true (\textit{left}) and absolute (\textit{right}) values.}\label{fig:coeffs}
\end{figure*}

Instead of applying the general perturbative framework covered in Section \ref{sec:Kato}, the summarized DMEs given in \tref{tab:DMEs} allow for a more concise and insightful calculation of the perturbational series up to order $n=9$. Moreover, since the individual orders $n$ share several common EMEs, the computational effort can be greatly reduced by first calculating all occurring EMEs and storing them for later access.\\
Figures \ref{fig:conv927} and \ref{fig:conv927rel} contain total perturbative energy corrections up to all orders $n=1,...,9$ for the Bose-Hubbard trimer, i.e. $M=3$, with system sizes of $N=9,27$. Since the first order correction is zero, the graphs for $n=1$ indicate the ground state energy of the unperturbed system described by $\widehat{H}_0$. \Fref{fig:conv927} contains absolute energy values for $n$-th order calculations as well as the result obtained from numerical diagonalization of the full Hamiltonian $\widehat{H}$ denoted by $E_d$. \footnote{The diagonalization procedure used is \texttt{scipy.linalg.eigh()} from \texttt{Scipy} version 1.14.1 together with \texttt{Python} version 3.12.} The convergence radii, determined using \eref{eq:convBH}, are indicated through vertical dashed lines. \\
It can be seen that for $\Omega/\kappa$ being outside the convergence radius, the present highest-order approximation for  $N=9$ still follows the course of $E_d$, whereas the one for $N=27$ quickly deviates from the corresponding numerical diagonalization result. However, for both system sizes, control parameter values within the estimated convergence radii coincide with regions where $E_d$ and curves for $n\geq 3$ are in accordance with each other - indicating a good convergence behavior of both series. \Fref{fig:conv927rel} contains the same energy calculations, however, scaled to the numerically computed ground state energy  $|E^{(n)} -E_d|/|E_d|$ - thus allowing for a better quantitative understanding of the energy corrections as well as a comparison of both system sizes. In general, with increasing order $n$, $E^{(n)}$ seems to approach $E_d$ - that is, at least for small $\Omega/\kappa$. For $N=9$, the seventh and eighth orders partly describe $E_d$ better than the ninth order does, even within the estimated convergence radius. For $N=27$, inside the estimated radius of convergence, only $n=7$ is slightly more accurate than $n=9$. This is a consequence of the series' alternating behavior demonstrated in \fref{fig:coeffs} by plotting the coefficients $c_n$ of the orders $n$ for different system sizes $N$ with both the true and absolute values. It can be seen that the coefficients exhibit an oscillating behavior with increasing amplitude for growing orders of $\Omega/\kappa$. The increase in amplitude can be seen in the right plot of \fref{fig:coeffs}: For larger systems, both the amplitude and its slope are higher. This gives reason to presume better convergence properties for smaller systems - a finding that is in accordance with \eref{eq:convBH}.

\section{Discussion}
Through the usage of the system's spatial symmetry and its spectral properties, as well as the hermiticity of its Hamiltonian, the standard Kato formalism can be reduced in terms of computational effort and structural simplicity. As seen in \tref{tab:numcontributions}, these simplifications lead to a reduction of terms required for a full computation of a perturbational order by almost a factor of two. Hence, for the determination of the ninth-order energy correction (while assuming convergence), one only needs to compute 18 different elementary matrix elements and combine all terms following the guide given in \tref{tab:DMEs}. For the two exemplary systems with $N=9,27$ considered here, even only sixth-order approximations sufficed in accuracy within estimated regions of convergence. For this, only six distinct elementary matrix elements would need to be calculated for the full computation. Moreover, the above-mentioned findings are inherently independent of the system size $M$ and the total number of particles $N$. As already described in \sref{sec:KatoBH}, the simplification scheme can be applied to a whole class of similar Bose-Hubbard systems, as long as the energy gap remains $1 \hbar \kappa$ and the perturbation only contains a single tunneling event. This also includes higher dimensional lattice geometries, such as hypercubic or triangular lattices for instance. As for the number of particles $N$, this solely influences the dimensionality of the Fock space and its elements which is given by \cite{kolovsky2016bosehubbardhamiltonian} 
\begin{eqnarray}
	\mathcal{D} = \frac{(N+M-1)!}{N!(L-1)!},
\end{eqnarray} 
i.e. the number of possible distributions of $N$ identical and indistinguishable particles onto $M$ lattice sites. For the trimer, this accounts to $\mathcal{D} = N(N+3)/2 +1$. While the simplification scheme itself does not scale with the system size, the numerical diagonalizations used as a reference for cross-checking do. Hence, the trimer with solely $\mathcal{D}\propto N^2$ and, thus, a diagonalization with reasonable computation effort serves as the optimal candidate for a minimal working example of our appraoch. \\
The degree of simplification achieved in this paper is partly owed to the reduction of $n$-th order contributing DMEs by explicitly excluding any first-order corrections, as described in \cite{eckardt2009processchainapproachhighorder,teichmann2009processchainapproachbosehubbard}. Both groups of authors use a figurative interpretation of the energy corrections: all $n$-th order contributions are regarded as sums over process chains that each are formed by a sequence of $n$ perturbation events that start and end in the unperturbed state $|a\rangle$. Each of these events within a sequence is then a matrix element of the perturbation operator $\langle \alpha |\widehat{V}|\beta \rangle$. With this, the structure of the perturbative corrections can be understood as sums over weighted classes of $n$-step closed pathways over the lattice. For the closed one-dimensional lattices of Bose-Hubbard rings, these $n$-step pathways are easy to identify as each lattice site only allows for two directions of travel. Closed pathways are then recognized by finding paths with ``effective''-steps (i.e. ignoring repetitive hoppings between two adjacent sites) that are multiples of $M$. If a path revisits its initial site multiple times, it can be split into independent sub-paths. This is formally equivalent to splitting a diagonal matrix element into elementary matrix elements. The hermiticity of the ring system Hamiltonians can be used to consider a path and its direction-reversed version as equal-valued - a finding that is equivalent to the reflection symmetries of the elementary matrix elements. A connection between the pathway interpretation and the analytic simplification of the elementary matrix elements in terms of the $k_i$ exponents using the unit spectral gap is yet to be found and may be covered in future publications. This might provide a still better and more thorough understanding of the application of the Kato formalism to Bose-Hubbard rings.\\
In this context, a figurative interpretation of the convergence criterion would also be desirable. \Eref{eq:convBH} already provides an analytical interpretation. Assuming Kato's criterion in \eref{eq:conv} to be applicable to Bose-Hubbard rings, the perturbation series converges and therefore delivers reasonable energy corrections if the ratio between the on-site interactions $\kappa$ and the tunneling frequency $\Omega$ is at least four times as big as the number of bosons. In Section \ref{sec:calc}, the calculated energy corrections give reason to trust this criterion: Here, in both exemplary systems, the higher-order approximations match the numerically aquired diagonalization procedure result. Outside the convergence radius, the diagonalization result (which does not need to follow any convergence criterion) veers off from the perturbative calculations - this is especially visible in the exemplary case of $N=27$. In this paper, only positive values for the control parameter $\Omega/\kappa$ were considered. As a convergence criterion is always two-sided, i.e. it considers the absolute value of the control parameter, this should also be applicable to comparable models with a positive tunneling contribution. \\
In general, the Kato formalism \cite{kato1949convergenceperturbationmethod,kato2013perturbationtheorylinear} appears to provide a more accessible approach to the structure of time-independent perturbation theory as formulated by Rayleigh and Schrödinger \cite{rayleigh1896theory,schrodinger1926quantisierungalseigenwertproblem}. The considerations shown in this present paper provide an example of how a formal approach to a specific type of quantum mechanical system can further reduce the complexity of these calculations and provide tools to gain more insight into perturbative mechanisms. By using system-specific properties in terms of symmetries and spectral characteristics, a formally expressed (recursive) sum can be traced back to a comparatively small set of irreducible elementary matrix elements whose values need to be known. An application to different (and potentially dimensionally higher) lattice geometries, see for instance \cite{teichmann2009processchainapproachbosehubbard} with hypercubic lattices, might turn out to both be very insightful and practical for approximating exact diagonalization procedures which for large systems can computationally become quite extensive.
\ack
I want to express my gratitude to Martin Holthaus for his guidance and support throughout the whole creation of this paper, especially for his feedback and concluding remarks on the manuscript. Further thanks goes to Thomas Löw for his helpful input during the early stages of code generation. I extend my gratitude to the University of Oldenburg for providing the financial resources and an academic environment conducive to research and development.
\section*{References}
\bibliographystyle{iopart-num}
\bibliography{sources.bib}
\appendix
\section{Hamiltonian for the Bose-Hubbard Trimer}\label{sec:appendix}
For the trimer, the Fock states in the $N$-particle sector of the Fock space $\mathcal{H}_N$ are uniquely defined by the occupation numbers of the three lattice sites $j,k,l$. Mathematically, the states can be expressed as product states of single-site wavefunctions $ |j\rangle, |k\rangle, |l \rangle$.
\begin{eqnarray}
	|j,k,l \rangle & = |j\rangle \otimes |k\rangle \otimes |l \rangle\\
	& = \frac{1}{\sqrt{\vphantom{x^2}j!~k!~l!}} \left (\hat{a_1}^\dag \right)^j ~\left (\hat{a_2}^\dag \right)^k  ~ \left (\hat{a_3}^\dag \right )^l  ~|\mathrm{vac} \rangle
\end{eqnarray}
Here, $\hat{a_i}^\dag$ is the creation operator for site $i$ ($i= 1,2,3$) and $|\mathrm{vac} \rangle$ describes the vacuum state. \cite{gallemi2015fragmentedcondensationbose}
With the help of the constraint $j+k+l = N = \mathrm{const.}$, each Fock state is defined by only two indices:
\begin{eqnarray}
	|j, k\rangle ~=~ |j,k , N-j-k\rangle 
\end{eqnarray}
The interaction terms in  $\widehat{H}$ act on a Fock state as follows:
\begin{eqnarray}
	\hat{a}^\dag_1 \hat{a}^\dag_1 \hat{a}_1 \hat{a}_1 ~ |j,k\rangle & = j(j-1)~ |j, k\rangle \\
	\hat{a}^\dag_2 \hat{a}^\dag_2 \hat{a}_2 \hat{a}_2 ~ |j,k\rangle & = k(k-1) ~|j, k\rangle \\
	\hat{a}^\dag_3 \hat{a}^\dag_3 \hat{a}_3 \hat{a}_3 ~ |j,k\rangle & = (N-j-k)~(N-j-k-1) ~|j, k\rangle ,
\end{eqnarray}
whereas the tunneling terms yield the following:
\begin{eqnarray}
	\hat{a}^\dag_1 \hat{a}_2 |j,k \rangle & = \sqrt{(j+1)k} ~|j+1,k-1 \rangle  \\
	\hat{a}^\dag_2 \hat{a}_1 |j,k \rangle & = \sqrt{j(k+1)} ~|j-1,k+1 \rangle \\[1em]
	\hat{a}^\dag_2 \hat{a}_3 |j,k \rangle & = \sqrt{(N-j-k)(k+1)} ~|j,k+1 \rangle\\
	\hat{a}^\dag_3 \hat{a}_2 |j,k \rangle & = \sqrt{(N-j-k+1)k}~ |j,k-1 \rangle \\[1em]
	\hat{a}^\dag_3 \hat{a}_1 |j,k \rangle & = \sqrt{j(N-k-j+1)} ~|j-1,k \rangle \\
	\hat{a}^\dag_1 \hat{a}_3 |j,k \rangle & = \sqrt{(j+1)(N-k-j)} ~|j+1,k \rangle .
\end{eqnarray}
Hence, the matrix elements of $\widehat{H}$ are \footnote{Adapted from \cite{chefles1996nearestneighbourlevelspacings}.}:
	\begin{eqnarray}
\langle ~ j',k'|\widehat{H}| j, k ~\rangle & = \frac{1}{2}\biggl [ 2 (j^2+k^2) + N^2 +2jk-N -2N(j+k)  \biggr ] \delta_{j,j'}\delta_{k,k'} \label{eq:H} \nonumber\\
& ~-\frac{\Omega}{\kappa} \biggl [ \sqrt{(j+1)~k} ~\delta_{j',j+1}~\delta_{k',k-1} \nonumber \\
& \qquad \quad+ \sqrt{j(k+1)}\delta_{j',j-1}\delta_{k',k+1} \nonumber \\
& \qquad \quad+ \sqrt{(N-j-k)(k+1)} ~\delta_{j,j'}~ \delta_{k',k+1} \nonumber \\
&\qquad  \quad+ \sqrt{(N-j-k+1)k}~ \delta_{j,j'}\delta_{k',k-1} \nonumber \\ 
&\qquad \quad + \sqrt{j(N-k-j+1)}~ \delta_{j',j-1} \delta_{k,k'} \nonumber \\
& \qquad\quad  + \sqrt{(j+1)(N-k-j)} ~\delta_{j',j+1} \delta_{k',k} \biggr ]. 
	\end{eqnarray}
The dimensionality of $\widehat{H}$, i.\,e.\ the total number of possible states, can be obtained by calculating the number of possibilities of partitioning $N$ particles into three groups:
\begin{eqnarray}
	\mathcal{D} := \mathrm{dim}\mathcal{H}_N = \sum_{j=0}^N N+1 - j = N~\frac{N+3}{2}+1 \label{eq:dimH}
\end{eqnarray} 
Therefore, for $N$ particles, $\widehat{H}$ is represented by a ${\mathcal{D}\times \mathcal{D}}$ square matrix, where each column or row index is mapped to a tuple $(j,k)$. The number of elements in $\hat{H}$ therefore increases with $N^4$. For the mapping, A. Chefles
\cite{chefles1996nearestneighbourlevelspacings} proposes a bijective transformation $b(j,k)$ which maps the two integer indices $(j,k)$ onto a single integer index $b$: 
\begin{eqnarray}
	&B: (j,k) \mapsto b , \quad j,k,b \in \mathbb{N}_0~, \quad j+k\leq N,\quad 0\leq b< d  \nonumber\\
	&B (j,k) = j\left(N-\frac{j-3}{2}\right) + k
\end{eqnarray}
\begin{table}[]
	\caption{Transformation of index $b$ into particle occupation numbers $j$,$k$, and $l$ for $N=3$ where dim$\mathcal{H}_3=10$.} \label{tab:bjk}
	\begin{indented}
		\item[]
		\begin{tabular}{ccccccccccc} \br
			b &0&1&2&3&4&5&6&7&8&9 \\ \cr\mr
			j &0&0&0&0&1&1&1&2&2&3 \\
			k &0&1&2&3&0&1&2&0&1&0 \\
			l &3&2&1&0&2&1&0&1&0&0 \\ \cr \br
		\end{tabular}
	\end{indented}
\end{table}\\
The inverse transformation can formally be expressed as:
\begin{eqnarray}
	B^{-1}:b & \mapsto (j,k), \quad j,k,b \in \mathbb{N}_0,~ j+k\leq N ~,\quad 0\leq b< d  \nonumber\\
	j &= \frac{2N+3}{2}-\sqrt{\left ( \frac{2N+3}{2}\right )^2-2b+2k}
\end{eqnarray}
For each index $b$, this equation only leads to a non-negative integer number $j$ for one specific (non-negative integer) $k$, therefore mapping each $b$ to a unique tuple $(j,k)$. \Tref{tab:bjk} shows the bijective mapping of all state tuples $(j,k)$ to indices $b$ for the example of $N=3$.

\end{document}